\documentclass[journal]{IEEEtran}
\ifCLASSINFOpdf
 
\else

\fi
\usepackage{amsmath,amssymb}
\makeatletter
\newsavebox\myboxA
\newsavebox\myboxB
\newlength\mylenA

\newcommand*\xoverline[2][0.75]{%
	\sbox{\myboxA}{$\m@th#2$}%
	\setbox\myboxB\null% Phantom box
	\ht\myboxB=\ht\myboxA%
	\dp\myboxB=\dp\myboxA%
	\wd\myboxB=#1\wd\myboxA% Scale phantom
	\sbox\myboxB{$\m@th\overline{\copy\myboxB}$}%  Overlined phantom
	\setlength\mylenA{\the\wd\myboxA}%   calc width diff
	\addtolength\mylenA{-\the\wd\myboxB}%
	\ifdim\wd\myboxB<\wd\myboxA%
	\rlap{\hskip 0.5\mylenA\usebox\myboxB}{\usebox\myboxA}%
	\else
	\hskip -0.5\mylenA\rlap{\usebox\myboxA}{\hskip 0.5\mylenA\usebox\myboxB}%
	\fi}
\makeatother
\usepackage{cite}
\usepackage{varioref}
\usepackage{amsmath}
\usepackage[usenames, dvipsnames]{color}

\usepackage{graphicx}
\usepackage[ruled]{algorithm2e}
\usepackage{threeparttable}
\usepackage{amsthm}
\usepackage{theoremref}
\usepackage{caption}
\makeatletter
\newcommand{\removelatexerror}{\let\@latex@error\@gobble}
\makeatother
\usepackage{amsmath}

\begin{document}

\title{Protocol design for energy efficient OLT transmitter in TWDM-PON guaranteeing SLA of up-stream and down-stream traffic}

\author{Sourav~Dutta, Dibbendu~Roy,~and~Goutam~Das% <-this % stops a space
\thanks{Sourav Dutta is with the Department
of Electronics and Electrical Communication Engineering, Indian institute of Technology Kharagpur,  Kharagpur,
India (e-mail: sourav.dutta.iitkgp@gmail.com).}% <-this % stops a space
\thanks{Dibbendu Roy and Goutam Das are with G. S. Sanyal School of Telecommunication, Indian Institute of Technology Kharagpur, Kharagpur, India (e-mail: dibbaroy@gmail.com, gdas@gssst.iitkgp.ernet.in).} % <-this % stops a space
}
\maketitle
\date{\vspace{-5ex}}
\vspace{-2cm}
\begin{abstract}
Environmental and economic concerns promote research on designing energy-efficient Time and Wavelength Division Multiplexed Ethernet Passive Optical Network (TWDM-EPON), which is the future extension to TDM-EPON. In TDM-EPON, a plethora of research is already present to achieve energy savings at Optical Network Units (ONUs) which can easily be applied for TWDM-EPON ONUs. However, TWDM-EPON provides an additional opportunity for saving energy at the Optical Line Terminal (OLT). All existing protocols have primarily been designed for saving energy at the OLT receivers. The protocols to save energy at the OLT receives depends only on the Up-Stream(US) traffic scheduling while its transmitter counterpart depends on both US and Down-Stream (DS) scheduling since the OLT transmits GATE message along with DS traffic. The US and DS scheduling have a basic difference. The MAC protocol doesn't allow scheduling of US traffic of an ONU after its REPORT arrival at multiple disjoint time slots. However, this restriction is absent for DS traffic and hence, the grant-size of an ONU can be partitioned and every part can be scheduled at different times. In this paper, we propose a method for saving energy at the OLT transmitters in TWDM-EPON while satisfying the SLAs. This includes a heuristic algorithm to partition the DS grant and schedule them. Through extensive simulations, we demonstrate that the proposed method provides a significant improvement in energy efficiency as compared to existing protocols (up to $\sim 45\%$).   
\end{abstract}

\begin{IEEEkeywords}
	EPON, TWDM-EPON, energy efficiency, Down-Stream scheduling, MAC protocol. 
\end{IEEEkeywords}
\IEEEpeerreviewmaketitle
\section{Introduction}\label{sec:intro}
Depletion of environmental impact and operational expenditure promotes research on designing energy-efficient Internet network especially the access network as it is the predominant source of Internet power consumption \cite{kani2013power}. The Ethernet Passive Optical Network (EPON) is one of the most widely accepted access networks which comprises an Optical Line Terminal (OLT), multiple Optical Network Units (ONUs), and single or multiple stages of Remote Nodes (RNs) \cite{ipact}. In EPON, the OLT and ONUs are the active components and hence, energy efficiency can be achieved by switching off transceivers of both the OLT and ONUs. In Time Division Multiplexed EPON (TDM-EPON), a single transceiver is used at the OLT for providing service to all ONUs and hence, it cannot be switched off for saving energy. Thus, in TDM-EPON, several protocols have been proposed for energy-efficient ONU design \cite{onu1,onu2,onu3,onu4,onu5,onu6}. However, to cope with the perpetual growth of Internet users, Time and Wavelength Division Multiplexed EPON (TWDM-EPON) is emerging to be the most promising access technology where multiple wavelengths (separate transceiver for every wavelength) are used for providing service to ONUs. The most popular TWDM-EPON architecture equips ONUs with tunable transceivers and RNs are realized with Power Splitters which provide full flexibility (any ONU can be scheduled to any wavelength) \cite{twdm_arch}. Thus, at low load, services can be provided by using fewer wavelengths while the transceivers, corresponding to remaining wavelengths, can be switched off for saving energy \cite{twdm1}. This provides an additional opportunity of achieving energy efficiency at the OLT in TWDM-EPON. However, in this case, packet delay increases and hence, protocols, to achieve energy efficiency, must ensure the SLAs.  
%The protocols that are proposed for saving energy at ONUs for TDM-EPON can be employed in TWDM-EPON as well with slide modification. 
Thus, in this paper, our focus is on designing protocol for saving energy at the OLT while ensuring the SLAs in a full-flexible TWDM-EPON.    

Energy efficiency at the OLT can be achieved by switching off both the transmitters and the receivers. Since the OLT receivers process only the Up-Stream (US) traffic, protocols for designing energy-efficient OLT receivers, consider only the US traffic. Several such protocols are already present in the literature \cite{twdm1,twdm2,twdm3,twdm4,twdm5,twdm6,twdm7,twdm8}. Among them, the fundamental principle of the protocols that are proposed in \cite{twdm1,twdm2,twdm3,twdm4,twdm5,twdm6,twdm7}, are the same. In all these protocols, only some of the available wavelengths, termed as "active wavelengths", are used for scheduling while the receivers, corresponding to the remaining wavelengths, termed as "inactive wavelengths", are switched off for saving energy. Thus, the primary objective of all these proposals is to minimize the number of active wavelengths. The process that is followed by the OLT to calculate the number of active wavelengths is termed as Dynamic Wavelength Allocation (DWA). The DWA is performed only at some predefined time instants (DWA boundaries) and in between two DWA boundaries, scheduling of all ONUs are performed only in the active wavelengths by following online \cite{twdm2} or offline \cite{twdm3,twdm4,twdm5,twdm6,twdm7} Dynamic Bandwidth Allocation (DBA) protocol. In \cite{twdm8}, we have proved that within a very large time interval $T$ ($T\rightarrow \infty$), the total duration, over which the OLT transceivers can be switched off (i.e., average energy efficiency), is inversely proportional to the number of voids ($M_v$) (idle periods between two consecutive data transmissions) and the number of REPORT messages (for receivers) or GATE messages (for transmitters) ($M_R$) present in this $T$ interval. Thus, energy efficiency at the OLT can be maximized by minimizing $M_v$ and $M_R$. We have also logically explained that the effect of minimization of $M_R$ is negligible. Thus, by employing the principle of minimization of the number of voids, in \cite{twdm8}, we have proposed a protocol,  namely EO-NoVM, for saving energy at OLT receivers and we have demonstrated that a significant improvement in energy efficiency can be achieved as compared to the conventional methodology of minimizing the number of active wavelengths.   

To save energy by switching off the OLT transmitters, one needs to consider the DS traffic. Usually, in a centralized network-paradigm like TWDM-PON, the US bandwidth allocation (MAC) is managed by employing a polling-based scheme where a control message (to poll) is sent along with the normal DS traffic. Although the US MAC protocol for TWDM-EPON has not been standardized, the authors of \cite{mac} have shown that the Multi-Point Control Protocol (MPCP) \cite{mpcp}, standardized US MAC protocol for TDM-EPON, can even be used for TWDM-EPON  with minor modifications. In this paper, we consider the same, as the US MAC protocol. In this protocol, the OLT polls every ONUs using GATE message where the wavelength information of both transmitter and receiver of the ONU along with the granted bandwidth are provided. After receiving the GATE message, an ONU tunes its transceiver to the informed wavelengths and up-streams the granted bandwidth.  Thereafter, a REPORT message is transmitted where the ONU informs its current queue length to the OLT which is then utilized for calculating the US grant-size and the time instant of the next GATE message transmission. Since the OLT transmitters send Down-Stream (DS) traffic along with the GATE message, consideration of both US and DS is required while saving energy at the OLT transmitters. 

To the best of our knowledge, there doesn't exist any protocol that can save energy at the OLT transmitters, except our previous proposal \cite{twdm8} where we have shown how EO-NoVM, can be employed for energy-efficient OLT transmitter design. However, the EO-NoVM protocol was primarily designed for scheduling the US traffic for saving energy at the OLT receivers. It is well known that US and DS traffic scheduling in EPON has a fundamental difference. In EPON, the OLT schedules the US traffic using GATE message where it specifies from where and for which duration the ONU transmits its US traffic. Thus, US traffic scheduling should be performed within a single time slot. However, this constraint is not present for DS traffic scheduling and hence, the DS grant can be partitioned and every part can be scheduled at multiple disjoint time slots. This allows taking a better decision which provides an additional opportunity of reducing the number of voids, resulting in an enhancement in energy efficiency. Further, this can provide more opportunity for satisfying SLA requirements. None of the existing protocols explore this opportunity. So, it is important to design a protocol to partition the DS traffic and to decide the time slots, at which they are scheduled. The protocol should be designed such that the number of voids gets minimized which finally leads to maximization of energy efficiency at the OLT transmitters and at the same time, the SLA requirements should be maintained.

In this paper, we design a heuristic algorithm, namely EOTx-NoVM (EOTx-NoVM- Energy-efficient OLT transmitter by Number of void minimization), to \textit{partition the DS grant and schedule them (i.e., DS traffic scheduling)}. The EOTx-NoVM protocol tries to minimize the number of voids (present at the OLT transmitters) while satisfying the SLA requirements. \textit{For minimizing the number of voids, the EOTx-NoVM algorithm takes a greedy approach i.e., every time the DS traffic is scheduled, the OLT tries to minimize as many voids as possible. If multiple such solutions are present then the OLT tries to schedule the DS grant such that it creates a better opportunity of minimizing the number of voids by future scheduling.}      
 % Thus, if multiple voids are present whose sizes are lower than the grant-size then US traffic cannot be scheduled in any of these voids. However, for DS traffic scheduling, this constraint is absent and hence,  DS traffic can be scheduled in multiple small voids which provides an additional opportunity of reducing the number of voids, resulting in an enhancement in energy efficiency. Further, this can provide more opportunity for satisfying SLA.
  %Also scheduling the DS data at the latest possible time instants would lead to queue build-up at OLT, which creates a better opportunity of void filling. Also, the SLA requirements are to be maintained which imposes restrictions on these opportunities.   
 % From the above discussion, it is now clear that designing a separate protocol, which will be primarily designed for achieving energy efficiency at OLT transmitters is required, unlike our previous proposal where the protocol, designed for energy-efficient OLT receivers (EO-NoVM protocol), is used for transmitters as well. 
%The major contribution of this paper is to design a protocol, namely EOTx-NoVM (EOTx-NoVM- Energy-efficient OLT transmitter by Number of void minimization) which tries to minimize the number of voids and schedule at the latest possible instants in order to maximize the achieved energy efficiency and at the same time the SLA requirements of both US and DS traffic are met. 
Through extensive simulations, we demonstrate that the proposed EOTx-NoVM protocol provides a significant improvement in energy efficiency (up to $\sim 45 \%$) when compared to our previously proposed protocol of \cite{twdm8}, which is the only existing protocol for saving energy at the OLT transmitters. Simulation results also depict that the probability of delay bound violation of the DS traffic is insignificant even up to a high traffic load.  

The rest of the paper is organized as follows. In Section II, the outline of our proposed mechanism for saving energy at the OLT in TWDM-EPON is explained. We provide a detail description of the proposed EOTx-NoVM protocol in Section III. Section IV presents the results and their brief discussion. The concluding statements are provided in Section V.  
\begin{table}[b]
	\begin{center}
		\caption{{Definition of notations}}
		\begin{tabular}{|c|p{7 cm}|}
			\hline
			\textbf{Notation} & \textbf{Description}\\
			\hline
			$N$ & Number of ONUs\\
			$T_{rtt}^k$ & Round trip time of $ONU_k$\\
			$T_t^{i,j}$ & Tuning time between wavelengths $i$ and $j$\\
			$T_g^k$ & Down-stream grant-size of $ONU_k$ (in time)\\
			$T_{rg}^k$ & Remaining grant of $ONU_k$\\
			$T_{sw}$ & Sleep-to-wake-up time for the OLT transmitters\\
			$D_{ub}^k$ & Delay bound of the down-stream traffic of $ONU_k$\\
			$W_p^k$ & Previously scheduled down-stream wavelength of $ONU_k$\\
			$t_{lb}^{k,j}$ & The minimum time instant, when the down-stream traffic of $ONU_k$ can be scheduled at wavelength $j$\\
			$t_{ub}^{k}$ & The maximum time instant, up to which  down-stream traffic of $ONU_k$ can be scheduled by satisfying the delay bound\\ 
			$t_m^j$ & The maximum time instant, up to which the DS traffic will be scheduled if the scheduling is performed in wavelength $j$\\
			$t_{lf}^j$ &Latest finish time of wavelength $j$\\
			$\mathbb{V}^j$ & Set of all voids at wavelength $j$\\
			$\mathbb{V}_f^j$ & Set of voids that get completely filled up if scheduling is performed in wavelength $j$\\
			$N_f^j$ & The number of voids that get completely filled up if scheduling is performed in wavelength $j$ i.e., $|\mathbb{V}_f^j|$\\
			$v_s^{i,j}$ & Start time of the $i^{th}$ void of $\mathbb{V}^j$\\
			$v_e^{i,j}$ & End time of the $i^{th}$ void of $\mathbb{V}^j$\\
			$v_{sz}^{i,j}$ & Size of the $i^{th}$ void of $\mathbb{V}^j$\\
			$v_v^{i,j}$ & $i^{th}$ element of set $\mathbb{V}_v^j$\\ 
			\hline 
		\end{tabular}
	\end{center}
\end{table}
\section{Framework of our proposed mechanism}\label{frame}
In this section, we first explain briefly the functionality of the OLT which is followed by the outline of the proposed EOTx-NoVM protocol.
\subsection{OLT functionality}
The OLT transmits both the DS traffic and GATE messages. The GATE message is scheduled by following the US traffic scheduling. Thus, maximization of the OLT energy savings  require designing a single  protocol for achieving energy efficiency both at transmitters and receivers by considering both US and DS traffic. However, for the sake of simplicity, we consider two different protocols run simultaneously for saving energy at the OLT receivers and transmitters. The scheduling protocol for energy-efficient OLT receives considers only the US traffic while its transmitter counter part considers only the DS traffic. Thus, in our proposed mechanism,  when a REPORT message arrives at the OLT from an ONU (say $ONU_k$), it first schedules the US traffic with the objective of maximizing the achieved energy efficiency at the OLT receivers while guaranteeing SLA of US traffic. To do this, the OLT follows our previously proposed EO-NoVM protocol \cite{twdm8} which calculates the GATE message transmission time ($t_{GATE}^k$) and the wavelength, at which $ONU_k$ transmits its up-stream data ($W_{tx}^k$), when it receive the GATE message.  We consider that the OLT informs an ONU about both the wavelengths at which its transmitter and receiver is scheduled through its own GATE message.   
% In our considered MAC protocol [], wavelengths, at which transceiver of an ONU will be tuned, are informed through the GATE message.
Clearly, the GATE message should be transmitted in the wavelength, at which the ONU was tuned by the previous GATE message, denoted as $W_{p}^k$. To ensure the GATE message transmission, the transmitter, corresponding to wavelength $W_{p}^k$, should be switched on at $t_{GATE}^k$ which modifies the void set of wavelength $W_{p}^k$ (i.e., set of all existing voids at wavelength $W_{p}^k$). In this modified void scenario, the DS traffic is then scheduled (i.e., at what wavelength and time intervals, the DS traffic will be transmitted) such that the energy savings at OLT transmitters get maximized and the SLA of DS traffic is guaranteed. Towards this target, the OLT first calculates the grant-size of the DS traffic of $ONU_k$ ($T_{g}^k$) by using any of the grant-sizing protocols \cite{ipact}. Thereafter, by using our proposed EOTx-NoVM protocol,  the OLT decides at which wavelength and time intervals, this $T_{g}^k$ amount of DS traffic will be down-streamed.  In this paper, we consider that the OLT follows the gated scheme for calculating $T_{g}^k$. Thus, in our proposed mechanism, the OLT follows the following steps at the reception of REPORT message from $ONU_k$:
\begin{itemize}
	\item The OLT runs our previously proposed EO-NoVM protocol for finding $t_{GATE}^k$ and $W_{tx}^k$.
	\item The void set of wavelength $W_{p}^k$ is then modified such that the GATE message transmission at wavelength $W_{p}^k$ and time $t_{GATE}^k$ is possible.
	\item Calculate $T_{g}^k$ by using gated scheme.
	\item Thereafter, the OLT runs the proposed EOTx-NoVM protocol for scheduling the DS traffic while satisfying its SLA.     
\end{itemize}    
 Next, we briefly provide an outline of the proposed EOTx-NoVM protocol.
 \subsection{Outline of EOTx-NoVM} \label{ssec:outline} 
  The proposed EOTx-NoVM protocol schedules the DS traffic such that  energy efficiency gets minimized while ensuring the SLA of the DS traffic. In this paper, we assume the delay bound is the only measure of SLA and a single class of traffic is present. Thus, the queuing delay of all DS traffic of $ONU_k$ is upper bounded by a fixed delay bound, denoted as $D_{ub}^k$. Satisfaction of this delay bound imposes a restriction on the voids at which the DS traffic of $ONU_k$ can be scheduled. 
  %In-fact, we show that the DS traffic can be scheduled in a wavelength, say wavelength $j$, within a time interval. 
  We now calculate the lower and the upper bound of this time interval and they are denoted by $t_{lb}^{k,j}$ and $t_{ub}^{k}$ respectively.
  \subsubsection{Calculation of $t_{lb}^{k,j}$} \label{sssc:lower}
  As mentioned above, in our considered MAC protocol, the OLT informs  at which wavelength an ONU need to tune its receiver through the GATE message. Thus, the DS traffic should be scheduled such that after reception of the GATE message, the ONU get enough time to tune its receiver. Let us denote the tunning time from wavelength $p$ to wavelength $j$ as $T_t^{p,j}$. If the GATE message for $ONU_k$ is transmitted at time $t_{GATE}^k$ and wavelength $W_p^k$ then the DS data at wavelength $j$ can only be scheduled after $t_{GATE}^k+T_t^{W_p^k,j}$.  It may also happen that the GATE message is sent before the completion of the previously scheduled DS data. In that case, if the ONU tune its receiver immediately after receiving the GATE message then the previously scheduled DS, which are not yet transmitted, is needed to the rescheduled. This increases the complexity of the DS scheduling protocol and hence, in this paper, we avoid such cases. Thus, the tunning should be performed after the completion of the previously scheduled DS data. It can be noted that the OLT needs to inform ONUs through GATE message when they perform tuning of their receivers along with the wavelength information. If $t_{lt}^{k,W_p^k}$ denotes the time instant up to which the DS traffic of $ONU_k$ has already been scheduled in wavelength $W_p^k$ then the DS traffic can only be scheduled in wavelength $j$ only after $t_{lt}^{k,W_p^k}+T_t^{W_p^k,j}$. Note that the OLT needs to inform ONUs through GATE message when they perform tuning of their receivers along with the wavelength information. Thus, $t_{lb}^{k,j}$ is given by eq. (\ref{lb}).
  \begin{align}\label{lb}
  t_{lb}^{k,j}=\max(t_{GATE}^k,t_{lt}^{k,W_p^k})+T_t^{W_p^k,j}
  \end{align}    
  Note that if $W_p^k=j$ then no tuning of the ONU receiver is required and hence, this lower bound on DS traffic scheduling, $t_{lb}^{k,j}$, is present only if $W_p^k\neq j$.  
  \subsubsection{Calculation of $t_{ub}^{k}$} \label{sssc:upper}
     The EOTx-NoVM protocol tries to schedule the DS traffic such that delay bound of every DS data is maintained, or in other words, every DS data of $ONU_k$ is transmitted within $D_{ub}^k$ duration after their arrival to the OLT. It can be noted that if the delay bound of the DS data that are at the front of the queue can be satisfied then it also guarantees the delay bound of every DS traffic. If the arrival time of the DS traffic of $ONU_k$ which is at the front of the queue is denoted by $t_{af}^k$ then $t_{ub}^{k}$ is given by eq. (\ref{ub}).
     \begin{align}\label{ub}
     t_{ub}^{k}=t_{af}^k+D_{ub}^{k}~~\forall j
     \end{align} 
     
     So, eq. (\ref{lb}) and eq. (\ref{ub}) provides the time interval over which the DS data of $ONU_k$ in wavelength $j$ can be scheduled. Now, we explain how the OLT schedules the DS traffic  within this time interval. 
     \subsubsection{Rules to schedule DS traffic}
     Now, we explain the time intervals and the wavelength at which the DS traffic of $ONU_k$ is scheduled in between $t_{lb}^{k,j}$ and $t_{ub}^{k}$ after its REPORT arrival. While calculating them, the OLT tries to maximize energy efficiency which can be achieved by minimizing the number of voids (as mentioned above). In our proposed protocol, we consider that after a REPORT arrival, the DS traffic is scheduled in only one wavelength. So, for every wavelength, the OLT first calculates the time intervals at which the DS traffic will be transmitted and they are calculated with the objective of minimizing the number of voids. Thereafter, the OLT selects the wavelength, for which the number of voids gets minimized.    
     
     %Scheduling of DS traffic is performed with the objective of minimizing the number of voids in order to maximize energy-efficiency (refer Section \ref{sec:intro}). In this paper, a heuristic approach is proposed for doing this.
     \subsubsection*{\textbf{Calculation of time intervals}}
      The number of voids get minimized if the DS traffic completely fills maximum number of voids. This reduces the total number of voids by the number of voids that are completely filled. So, for every wavelength, the OLT first calculates the number of voids that can be completely filled-up ($N_f^j$) if the DS traffic is scheduled in it. To do this, the OLT first sort the voids corresponding to a wavelength in ascending order based on their size and goes through the sorted void set until the total size of the traversed voids is more than the grant-size (in time), $T_{g}^k$. Clearly, there doesn't exist any other void-fill process that can increase the value of $N_f^j$. It may occur some portion of $T_{g}^k$ doesn't get allocation by this process and this remaining grant-size (in time) is denoted as $T_{rg}^k$. 
    
     Note that the remaining grant ($T_{rg}^k$), cannot fill any of the unoccupied voids completely. In this case, the number of voids can be minimized in the following ways: 
     \begin{itemize}
     	\item Inter-ONU clubbing: As the remaining grant cannot fill any of the unoccupied voids completely the number of voids cannot be reduced further. In this case, the OLT tries not to increase the number of void. This is possible if the DS traffic is scheduled such that it gets clubbed with the DS traffic of other ONUs which is named as inter-ONU clubbing.
     	\item Creation of future Inter-ONU clubbing opportunity: If the DS traffic of an ONU (say $ONU_i$) is scheduled to the latest possible time instant then maximum number ONUs gets opportunity (in future) to club their DS data with the scheduled DS traffic of $ONU_i$. Thus, scheduling DS traffic to the latest possible time increases the opportunity of inter-ONU clubbing for future scheduling, causing an increment of energy efficiency. 
     	%Schedule the DS at the latest possible time instant which allows queue build-up. 
     \end{itemize}
 In this paper, inter-ONU clubbing is given more priority over scheduling to the latest possible time instant like the previously proposed EO-NoVM protocol.   
The remaining grant is scheduled in the following ways:
\begin{itemize}
	\item Try to schedule the remaining grant in the unoccupied voids or after the latest finish time such that inter-ONU clubbing is possible.
	\item If multiple such options of inter-ONU clubbing is present then schedule as late as possible.
	\item If inter-ONU clubbing is not possible then also scheduling is performed at the latest possible time instant. Note that in this case, an extra void will be created and hence, the value of $N_f^j$ is reduced by one. 
	\item It may also happen that the remaining grant can not be scheduled in between $t_{lb}^{k,j}$ and $t_{ub}^{k}$. In this case, if scheduling is performed in wavelength $j$ then there is a possibility of violation of SLA and the OLT tries to avoid scheduling in this wavelength.
\end{itemize}

\subsubsection*{\textbf{Selection of scheduled wavelength}} 
Till now, we have found the wavelengths at which scheduling can be performed in between $t_{lb}^{k,j}$ and $t_{ub}^{k}$, termed as valid wavelengths. For every valid wavelength, we have also calculated the time intervals at which the DS traffic of $ONU_k$ will be scheduled in the proposed EOTx-NoVM protocol. If no valid wavelengths are present then violation of SLA of DS traffic may happen. In this case, the OLT schedules DS traffic such that its transmission finish to the earliest time instant and the scheduled wavelength is that wavelength where this time instant is minimum. The next step is to decide the wavelength at which the DS data will be transmitted if a valid wavelength is present.   

As mentioned above, for every valid wavelengths, the OLT calculates the number of voids that can be reduced by scheduling the DS traffic  to that wavelength (i.e. $N_f^j$ $\forall j$). The other way to reduce the number of voids is by scheduling as late as possible. Thus, for every wavelength, the OLT also calculates the maximum time up to which the DS traffic is scheduled in that wavelength which is denoted as $t_{m}^j$ for wavelength $j$. Considering the values of $N_f^j$  and $t_{m}^j$, the OLT selects the wavelength in the following manner:
\begin{itemize}
	\item Select the wavelength, at which $N_f^j$ is maximum.
	\item If multiple maxima exist then select the wavelength at which $t_{m}^j$ is maximum. 
\end{itemize} 
Next, we explain our proposed EOTx-NoVM protocol in detail.
\begin{figure}[t]
	\centering
	\includegraphics[scale=.4]{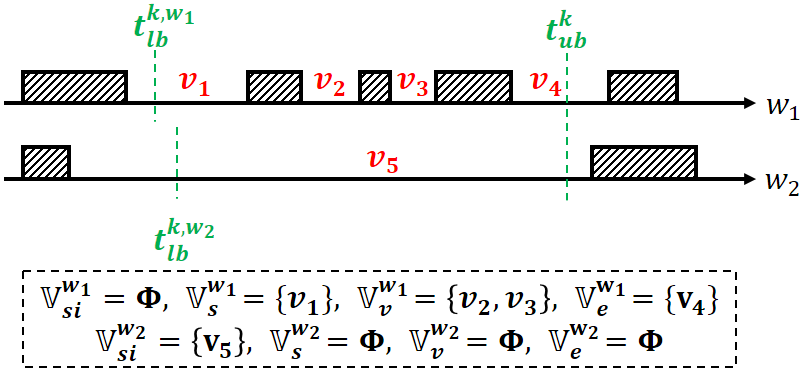}
	\caption{Types of voids}
	\label{void_type}
\end{figure} 

\section{EOTx-NoVM Protocol}   
\begin{figure*}[t]
	\centering
	\includegraphics[scale=.7]{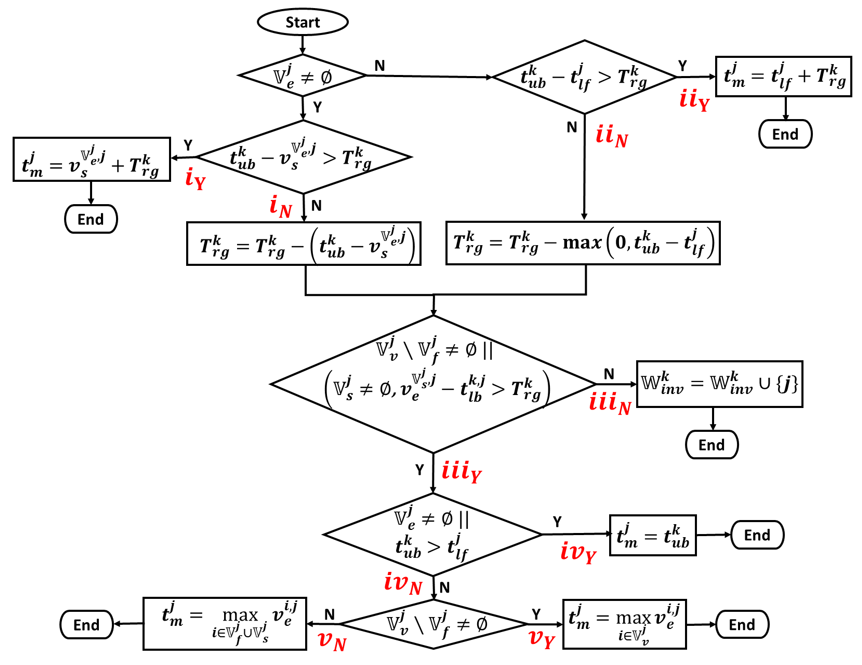}
	\caption{Scheduling of remaining grant and calculation of $t_m^ j$}
	\label{algo2}
\end{figure*}

The proposed EOTx-NoVM Protocol is employed to partition and schedule DS traffic with the objective of minimizing the number of voids while maintaining the SLA of DS traffic. To do so, for every wavelengths, the OLT first calculates the time intervals at which the DS traffic for an ONU are transmitted if scheduling is performed in that wavelength (refer Section \ref{frame}) and calculates two parameters: the number of voids that can reduced ($N_f^j$) and the maximum time up to which the DS traffic is scheduled ($t_{m}^j$). These two parameters are then used for calculating the DS wavelength by following the rules as discussed above. 
\begin{figure*}[t]
	\centering
	\includegraphics[scale=.73]{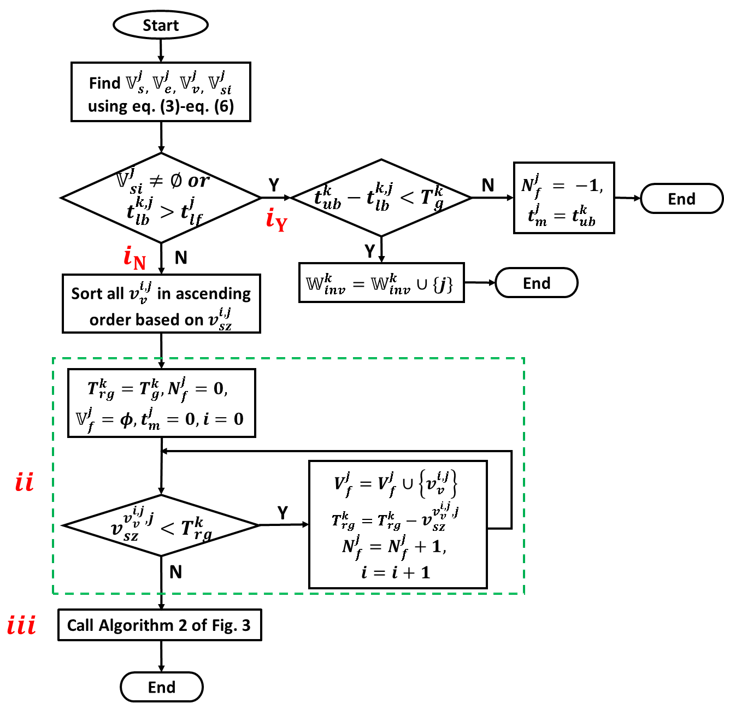}
	\caption{Finding completely filled void}
	\label{algo1}
\end{figure*}
Scheduling of the DS traffic can be performed either within voids or after the latest finish time. Let us denote the set of all wavelengths, the latest finish time of wavelength $j$ and the set of all voids at wavelength $j$ by $\mathbb{W}$, $t_{lf}^j$ and $\mathbb{V}^j$ respectively. Let the $i^{th}$ element of set $\mathbb{V}^j$ is $v^{i,j}$ which is a vector, $[v_s^{i,j}, ~v_e^{i,j}, ~v_{sz}^{i,j}]$, where $v_s^{i,j}$, $v_e^{i,j}$ and $v_{sz}^{i,j}$ denote the start-time, the end-time and the size of void $v^{i,j}$. Clearly, $v_{sz}^{i,j}=v_{e}^{i,j}-v_{s}^{i,j}$. Further, guaranteeing delay bound of DS traffic requires that scheduling of  $ONU_k$ at wavelength $j$ should be performed in between time instants $t_{lb}^{k,j}$ and $t_{ub}^{k}$ (refer Section \ref{sssc:lower} and \ref{sssc:upper}). Now, the start-time of a void (say $i$ i.e., $v_s^{i,j}$) at which scheduling can be performed can be either more or less than $t_{lb}^{k,j}$. Similarly,  $v_e^{i,j}$ can be either more or less than $t_{ub}^{k}$. Note that scheduling, in a void (say $v^ {i,j}$), can only be performed  if $t_{ub}^k>v_s^{k,j}$ and $t_{lb}^{k,j}<v_e^{k,j}$. Thus, a void, $v^{i,j}$, can be of four different types: $\mathbb{V}^j_s$, $\mathbb{V}^j_e$, $\mathbb{V}^j_v$, and $\mathbb{V}^j_{si}$ respectively and they are given by eq. (\ref{vs})-eq. (\ref{vsi}). This can also be seen from fig. \ref{void_type} where we consider two wavelengths ($w_1$ and $w_2$) are present. At $w_1$, voids $v_1$, $v_2$, $v_3$, and $v_4$ and at $w_2$, void $v_5$ are present.  
\begin{align}\label{vs}
\mathbb{V}^j_s=\{v^{i,j}| v_s^{i,j}<t_{lb}^{k,j}, v_e^{i,j}\leq t_{ub}^{k}~~\forall v^{i,j}\in V^j\}\\\label{ve}
\mathbb{V}^j_v=\{v^{i,j}| v_s^{i,j}\geq t_{lb}^{k,j}, v_e^{i,j}\leq t_{ub}^{k}~~\forall v^{i,j}\in V^j\}\\\label{vv}
\mathbb{V}^j_e=\{v^{i,j}| v_s^{i,j}\geq t_{lb}^{k,j}, v_e^{i,j}>t_{ub}^{k}~~\forall v^{i,j}\in V^j\}\\\label{vsi}
\mathbb{V}^j_{si}=\{v^{i,j}| v_s^{i,j}<t_{lb}^{k,j}, v_e^{i,j}>t_{ub}^{k}~~\forall v^{i,j}\in V^j\}
\end{align}
It is quite evident that these four sets satisfy the following properties:
\begin{itemize}
	\item All $\mathbb{V}^j_s$, $\mathbb{V}^j_e$, and $\mathbb{V}^j_{si}$ can contain at most one element while $\mathbb{V}^j_v$ can contain multiple elements (as seen at wavelength $w_1$ of Fig. \ref{void_type}).
	\item If $\mathbb{V}^j_{si}\neq \Phi$ then $\mathbb{V}^j_s=\mathbb{V}^j_e=\mathbb{V}^j_v=\Phi$ (as seen at wavelength $w_2$ of Fig. \ref{void_type}).
	\item If $\mathbb{V}^j_{si}\neq \Phi$ then scheduling in wavelength $j$ must create an extra void.
	\item If a void, $i$ can be completely fill-up by DS scheduling while satisfying delay bound then $i\in \mathbb{V}^j_v$.
\end{itemize}
Next, we describe the values of $N_f^j$ and $t_m^j$ for different cases with the help of Fig. \ref{algo1} and Fig. \ref{algo2} respectively where we have provided the flowchart of our proposed EOTx-NoVM algorithm. 
\subsection{Calculation of $N_f^j$}
As discussed above, if $\mathbb{V}^j_{si}\neq \Phi$ then scheduling of DS data at wavelength $j$ must create an extra void. The same argument holds true if $t_{lb}^{k,j}>t_{lf}^j$. Thus, for both of these two cases, $N_f^j=-1$ as seen from $i_Y$ of Fig. \ref{algo1}. 
%The scheduling should be performed with the objective of maximizing intra-ONU clubbing i.e. DS traffic is scheduled as late as possible while satisfying the delay bound. Thus, the DS traffic is scheduled such that the DS transmission terminates at time, $t_{ub}^j$ and hence, $t_m^j=t_{ub}^j$. 
For all other cases, it is always possible to schedule DS traffic without creating an extra void. The OLT tries to schedule DS traffic such that maximum number of voids get completely filled-up. At the same time, scheduling is performed to the latest possible time. This is a combinatorial optimization problem. Here, we propose a heuristic approach for solving this problem. Note that the voids of $\mathbb{V}_v^j$ can only be completely filled-up. Let us denote the $i^{th}$ element of $\mathbb{V}_v^j$ as $v_v^{i,j}$. In our proposed approach:
\begin{itemize}
	\item All voids of $\mathbb{V}_v^j$ (i.e., $v_v^{i,j}$ $\forall i$) are sort in ascending order based on their sizes as seen from $i_N$ of Fig. \ref{algo1}. 
	\item Voids are filled-up one after another from the beginning until all voids are completely filled-up or no further voids can be completely filled-up as seen at $ii$ of Fig. \ref{algo1}. Clearly, some remaining grant will be present after this step, which is termed as remaining grant and it is denoted by $T_{rg}^k$.
	\item Schedule the remaining grant, $T_{rg}^k$, using Algorithm 2 (refer Fig. \ref{algo2}) as seen at $iii$ of Fig. \ref{algo1}. .  
\end{itemize}
   It is evident that there doesn't exist any other allocation which can fill more number of voids completely as compared to our proposed approach. 
   %Let, us denote the set of all voids that get completely filled-up and the remaining portion of the grant-size of $ONU_k$ that is yet to be scheduled by $\mathbb{V}_f^j$ and $T_{rg}^k$ respectively. 
   Note that, $T_{rg}^k$ can be scheduled at wavelength $j$ in  $\mathbb{V}^j_s$, $\mathbb{V}^j_e$, $\mathbb{V}^j_v\setminus\mathbb{V}_f^j$ and after the latest finish time without creating extra void. Thus, in this case, $N_f^j=|\mathbb{V}_f^j|$ where $|\mathbb{V}_f^j|$ denotes cardinality of set $\mathbb{V}_f^j$. If $\mathbb{V}_v^j=\Phi$ then $N_f^j=0$ and $T_{rg}^ k=T_g^ k$ where $T_g^k$ is the DS grant-size of $ONU_k$. We now explain with the help of Fig. \ref{algo2}, how to schedule the remaining grant, $T_{rg}^k$ and the value of $t_m^j$ for all cases.
 \subsection{Remaining grant scheduling and calculation of $t_m^ j$}
 As discussed above, if $\mathbb{V}^j_{si}\neq \Phi$ or $t_{lb}^{k,j}>t_{lf}^j$ then none of the voids can be completely filled-up and inter-ONU clubbing is also not possible. Thus, scheduling of the DS traffic should be performed as late as possible. In this case, scheduling can be performed at any time in between $t_{lb}^ {k,j}$ and $t_{ub}^{k}$. Thus, if $t_{ub}^{k}-t_{lb}^{k,j}\leq T_g^k$, then scheduling in wavelength $j$ may cause violation of delay bound. So, include $j$ in set $\mathbb{W}_{inv}^k$ where $\mathbb{W}_{inv}^k$ denotes the set of wavelengths at which delay bound may gets violated. In other cases, scheduling is performed such that DS transmission terminates at $t_{ub}^{k}$ and hence, $t_m^j=t_{ub}^{k}$ as seen at $i_Y$ of Fig. \ref{algo1}.
 
 As mentioned above, in all other cases, inter-ONU clubbing is possible. Note that the remaining grant, $T_{rg}^k$, can be scheduled at wavelength $j$ in  $\mathbb{V}^j_s$, $\mathbb{V}^j_e$, $\mathbb{V}^j_v\setminus\mathbb{V}_f^j$ and after the latest finish time. Scheduling is performed as late as possible and at the same time, inter-ONU clubbing should be ensured. It can be noted that if $\mathbb{V}^j_e\neq \Phi$ then $t_{ub}^{k}<v_e^{\mathbb{V}_e^j,j}\implies t_{ub}^{k}<t_{lf}^j$. Hence, if  $\mathbb{V}^j_e\neq \Phi$ scheduling cannot be performed after the latest finish time. Similarly, if $t_{ub}^{k}>t_{lf}^j$ i.e., scheduling can be performed after the latest finish time then $\mathbb{V}^j_e= \Phi$. Clearly, if scheduling either in $\mathbb{V}^j_e$ or after the latest finish time is possible then it create maximum opportunity of inter-ONU clubbing for future ONUs. Hence, the OLT first tries to schedule the remaining grant either in $\mathbb{V}^j_e$ or after the latest finish time. 
 \subsubsection*{$\mathbf{\mathbb{V}^j_e\neq \Phi}$ \textbf{case}}
 Note that scheduling of DS traffic in $\mathbb{V}^j_e$, by satisfying the delay bound, can be performed in between $v_s^{\mathbb{V}_e^j,j}$ and $t_{ub}^{k}$. Thus, if $t_{ub}^{k}-v_s^{\mathbb{V}_e^j,j}>T_{rg}^k$ then remaining grant can be completely scheduled in void, $\mathbb{V}^j_e$ and the scheduling interval of it is in between $v_s^{\mathbb{V}_e^j,j}$ and $v_s^{\mathbb{V}_e^j,j}+T_{rg}^k$. Thus, in this case, $t_m^j$ is given by eq. (\ref{vephi}) as seen from $i_Y$ of Fig. \ref{algo2}.
 \begin{align}\label{vephi}
 t_m^j=v_s^{\mathbb{V}_e^j,j}+T_{rg}^k
 \end{align}
% Otherwise, remaining grant occupy the entire time interval in between $v_s^{\mathbb{V}_e^j,j}$ and $t_{ub}^{k,j}$ and hence, $t_m^j$ is given by eq. (\ref{vephi1}).
% \begin{align}\label{vephi1}
% t_m^j=t_{ub}^{k,j}
% \end{align}
  Otherwise, remaining grant must occupy the entire time interval in between $v_s^{\mathbb{V}_e^j,j}$ and $t_{ub}^{k}$ of $\mathbb{V}_e^j$. Clearly, in this case, $T_{rg}^k-(t_{ub}^{k}-v_s^{\mathbb{V}_e^j,j})$ amount of bandwidth is yet to be scheduled. Thus, remaining grant is updated by eq. (\ref{rg}) as seen from $i_N$ of Fig. \ref{algo2}.
  \begin{align}\label{rg}
  T_{rg}^k=T_{rg}^k-(t_{ub}^{k}-v_s^{\mathbb{V}_e^j,j})
  \end{align}
  \subsubsection*{$\mathbf{t_{ub}^{k}>t_{lf}^j}$ \textbf{case}}
  If $t_{ub}^{k}>t_{lf}^j$ then $\mathbb{V}^j_e=\Phi$ and the remaining grant can be scheduled after the latest finish time of wavelength $j$ within the time interval $t_{lf}^j$ and $t_{ub}^{k}$. Thus, if $t_{ub}^{k}-t_{lf}^j>T_{rg}^j$ then entire $T_{rg}^j$ is scheduled (i.e. $T_{rg}^k$ become zero) in between the time interval $t_{lf}^j$ and $t_{lf}^j+T_{rg}^j$ and in this case, the value of $t_m^j$ is given by eq. (\ref{vephi2}) as seen from $ii_Y$ of Fig. \ref{algo2}.
  \begin{align}\label{vephi2}
  t_m^j=t_{lf}^j+T_{rg}^j
  \end{align}
  Otherwise, if $t_{ub}^{k}-t_{lf}^j<T_{rg}^j$ then entire time interval between $t_{lf}^j$ and $t_{ub}^{k}$, must be occupied by remaining grant and the remaining grant will be updated by eq. (\ref{rg1}) as seen from $ii_N$ of Fig. \ref{algo2}.
% \begin{align}\label{vephi3}
% t_m^j=t_{ub}^{k,j}
% \end{align} 
 \begin{align}\label{rg1}
 T_{rg}^k=T_{rg}^k-t_{ub}^{k}+t_{lf}^j
 \end{align}
 
 After scheduling in $\mathbb{V}^j_e$ or after the latest finish time, if some portion of the grant-size is yet to be scheduled (i.e. $T_{rg}^k> 0$) then the remaining grant can be scheduled in a void of $\mathbb{V}_v^j\setminus \mathbb{V}_f^j$ or  the void $\mathbb{V}_s^j$. Note that, if there exists an void $i\in \mathbb{V}_v^j\setminus \mathbb{V}_f^j$ such that $v_e^{i,j}-v_s^{i,j}<T_{rg}^k$ then it must be included in set, $\mathbb{V}_f^j$. Thus, $v_e^{i,j}-v_s^{i,j}>T_{rg}^k$ $\forall i \in \mathbb{V}_v^j\setminus \mathbb{V}_f^j$. Hence, if $\mathbb{V}_v^j\setminus \mathbb{V}_f^j\neq \Phi$ then remaining grant can be completely scheduled in a void of $\mathbb{V}_v^j\setminus \mathbb{V}_f^j$.  
 The other possibility to perform scheduling of the remaining grant is in the void $\mathbb{V}_s^j$ where scheduling can be performed in between $t_{lb}^{k,j}$ and $v_e^{\mathbb{V}_s^j,j}$. Thus, if $v_e^{\mathbb{V}_s^j,j}-t_{lb}^{k,j}>T_{rg}^k$ then the remaining grant can be completely scheduled in the void $\mathbb{V}_s^j$ and otherwise, there is a possibility of violation of delay bound if the scheduling is performed in wavelength $j$ and hence, $j$ should be included in set $\mathbb{W}_{inv}^k$. Therefore, 
 $$  \mathbb{W}_{inv}^k=\mathbb{W}_{inv}^k\cup \{j\}~~\text{if}~~  \mathbb{V}_v^j\setminus \mathbb{V}_f^j= \Phi \text{ and } v_e^{\mathbb{V}_s^j,j}-t_{lb}^{k,j}<T_{rg}^k$$ which can be  seen from $iii_N$ of Fig. \ref{algo2}. 

Otherwise (refer $iii_Y$ of Fig. \ref{algo2}), delay bound of the DS traffic can always be satisfied. Note that in this case, if $\mathbb{V}_e^ j\neq \Phi$ then remaining grant must be scheduled in void $\mathbb{V}_e^ j$ up to $t_{ub}^{k}$ and some portion in a void of $\mathbb{V}_v^j\setminus \mathbb{V}_f^j$ or $\mathbb{V}_s^j$. Clearly, in this case, $t_m^ j=t_{ub}^{k}$.   
Similarly, if $t_{ub}^ {k}>t_{lf}^j$ then also $t_m^ j=t_{ub}^{k}$. This two cases can be seen from $iv_Y$ of Fig. \ref{algo2}.  
If $\mathbb{V}_e^ j=\Phi$ or $t_{ub}^ {k}<t_{lf}^j$ (refer $iv_N$ of Fig. \ref{algo2}) then remaining grant can only be scheduled in a void of $\mathbb{V}_v^j\setminus \mathbb{V}_f^j$ or $\mathbb{V}_s^j$. We now find the value of $t_m^ j$ for these two cases.  
 \subsubsection*{$\mathbf{\mathbb{V}_v^j\setminus \mathbb{V}_f^j\neq \Phi}$ \textbf{case}}
  Note that scheduling anywhere in a void of $\mathbb{V}_v^j\setminus \mathbb{V}_f^j$ creates more opportunity of inter-ONU clubbing for future ONUs as compared to perform scheduling in $\mathbb{V}_s^j$.
   Thus, scheduling in a void of  $\mathbb{V}_v^j\setminus \mathbb{V}_f^j\neq \Phi$ is prioritized as compared to scheduling in $\mathbb{V}_s^j$. If multiple voids are present in $\mathbb{V}_v^j\setminus \mathbb{V}_f^j$ then the OLT chooses the void $i_r$, for which the end-time is maximum i.e., $i_r=\arg \max\limits_{i\in \mathbb{V}_v^j\setminus \mathbb{V}_f^j} v_e^{i,j}$ and the remaining grant is scheduled such that the data transmission ends at the end-time of the void. 
   %As mentioned above, if $\mathbb{V}^j_e\neq \Phi$ or $t_{ub}^{k,j}>t_{lf}^j$ then scheduling of remaining must be performed in $\mathbb{V}^j_e$ or after the latest finish time and hence, $$t_m^j=t_{ub}^{k,j}$$ which is the highest possible value of $t_m^j$ as seen from $iv_Y$ of Fig. \ref{algo2}.
   In this case, the value of $t_m^j$ is either the maximum value of end-time of the voids that are completely filled-up (i.e. $\max\limits_{i\in \mathbb{V}_f^j} v_e^{i,j}$) or the end-time of the void at which the remaining grant is scheduled (i.e., $v_e^{i_r,j}$). Since  $i_r=\arg \max\limits_{i\in \mathbb{V}_v^j\setminus \mathbb{V}_f^j} v_e^{i,j}$, the value of $t_m^j$ is given by eq. (\ref{rg2}) as seen from $v_Y$ of Fig. \ref{algo2}.
   \begin{align}\label{rg2}
   t_m^j=\max\Big(\max\limits_{i\in \mathbb{V}_f^j} v_e^{i,j},v_e^{i_r,j}\Big)=
   \max\limits_{i\in \mathbb{V}_v^j} v_e^{i,j}
   \end{align} 
   Now, we find the value of $t_m^j$ for the case  $\mathbb{V}_v^j\setminus \mathbb{V}_f^j= \Phi$. 
    \subsubsection*{$\mathbf{\mathbb{V}_v^j\setminus \mathbb{V}_f^j= \Phi}$ \textbf{case}}
    In this case, only option to schedule the remaining grant is in the void, $\mathbb{V}_s^j$ where scheduling can be performed in between $t_{lb}^{k,j}$ and $v_e^{\mathbb{V}_s^j,j}$.
    % As mentioned above, if $\mathbb{V}_s^j= \Phi$ or $\mathbb{V}_s^j\neq \Phi$ and $v_e^{\mathbb{V}_s^j,j}-t_{lb}^{k,j}<T_{rg}^k$ then scheduling in wavelength $j$ may cause violation of delay bound of some DS traffic and hence, $j$ is included in set, $\mathbb{W}_{inv}$.
     Scheduling of remaining grant is performed to the latest possible time instant and hence, remaining grant is scheduled such that the DS transmission ends exactly at $v_e^{\mathbb{V}_s^j,j}$. Thus, in this case, $t_m^j$ is either maximum end-time of all voids of $\mathbb{V}_f^j$ or the end-time of $\mathbb{V}_s^j$ and hence, $t_m^j$ is given by  eq. (\ref{rg3}) as seen from $v_N$ of Fig. \ref{algo2}. 
       \begin{align}\label{rg3}
    t_m^j=\max\Big(\max\limits_{i\in \mathbb{V}_f^j} v_e^{i,j},v_e^{\mathbb{V}_s^j,j}\Big)
    \end{align}
    \begin{table}[b]
    	\begin{center}
    		\caption{{Parameters used in simulations}}
    		\begin{tabular}{|c|c|}
    			\hline
    			\textbf{Parameter} & \textbf{value}\\
    			\hline
    			Bandwidth of the feeder fiber & 1 Gbps\\
    			Bandwidth of the access fiber & 100 Mbps\\
    			Size of the GATE message & 64 Bytes\\
    			Size of the REPORT message & 64 Bytes\\
    			GATE processing time & 35 ns\\
    			Guard duration & 2 $\mu$s\\
    			$T_t$ & 1 $\mu$s\\	
    			\hline 
    		\end{tabular}
    	\end{center}
    \end{table}
    \subsection{Selection of scheduled wavelength}
    Till now we have found the time intervals at which the DS traffic of $ONU_k$ will be scheduled to maximize the energy efficiency if scheduling is performed at wavelength $j$ ($\forall j\in \mathbb{W}$). Our next step is to select the scheduled wavelength. While selecting the wavelength, the OLT must avoid any possibility of violation of delay bound for the DS traffic. As discussed above, if possibility of violation of delay bound is present for a wavelength then that wavelength is included in set $\mathbb{W}_{inv}$. Thus, wavelength selection is performed from the set $\mathbb{W}\setminus\mathbb{W}_{inv}$. If $\mathbb{W}\setminus\mathbb{W}_{inv}=\Phi$ then violation of delay bound of DS traffic may occur. Thus, in this case, the objective is to minimize the possibility of delay bound violation. Therefore, the DS traffic will be scheduled such that the DS transmission ends as early as possible and scheduled wavelength is that one at which termination of DS transmission occur first. Otherwise, scheduling can be performed in wavelengths of $\mathbb{W}\setminus\mathbb{W}_{inv}$ and the delay bound is met to be satisfied. In this case, the scheduled wavelength is selected with the objective of 
    \begin{itemize}
    	\item minimizing the number of voids (i.e. maximize $|\mathbb{V}_f^j|$) 
    	\item maximizing opportunity of inter-ONU clubbing for future ONUs (i.e maximize $t_m^j$)
    \end{itemize}      
In this paper, we prioritize the maximization of $|\mathbb{V}_f^j|$ over maximization of $t_m^j$. Thus, the OLT first find the wavelengths at which the value of $|\mathbb{V}_f^j|$ is maximum (i.e. $\mathbb{W}_f=\arg\max\limits_{j\in \mathbb{W}\setminus\mathbb{W}_{inv}} |\mathbb{V}_f^j|$). If $\mathbb{W}_f$ has multiple wavelengths then select the wavelength ($\in \mathbb{W}_f$) for which $t_m^j$ is maximum (i.e. $\mathbb{W}_m=\arg\max\limits_{j\in \mathbb{W}_f} t_m^j$). If set $\mathbb{W}_m$ has multiple wavelengths then the OLT chooses a wavelength randomly from $\mathbb{W}_m$.
\begin{figure}[t]
	\centering
	\includegraphics[scale=.7]{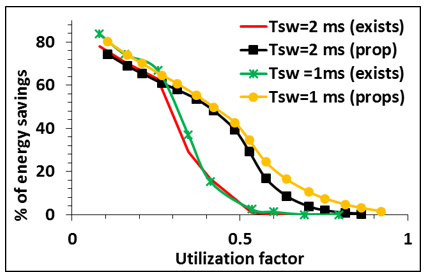}
	\caption{Comparison with the previously proposed EO-NoVM protocol.}
	\label{compari}
\end{figure}
\begin{figure*}
	\centering
	\includegraphics[scale=.7]{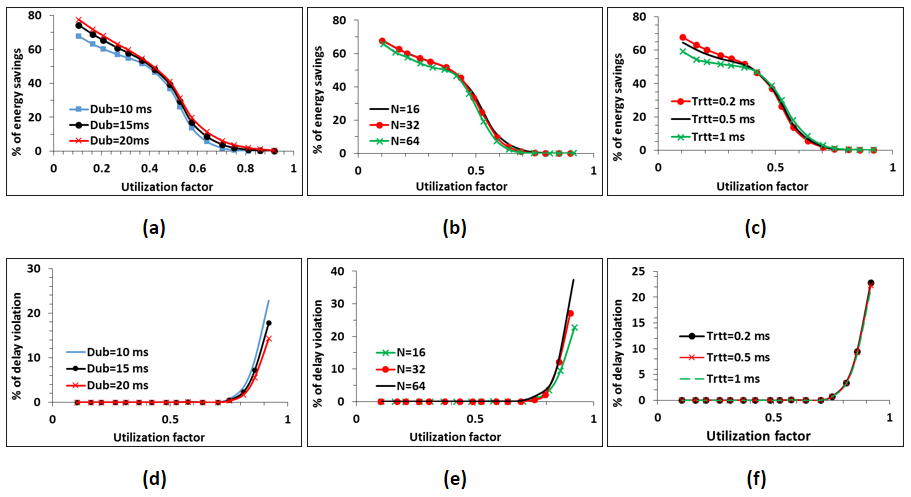}
	\caption{Energy efficiency for different value of (a) $D_{ub}^k$, (b) $N$ (c) $T_{rtt}^k$ and percentage of delay bound violation  for different value of (d) $D_{ub}^k$, (e) $N$ (f) $T_{rtt}^k$ of the proposed EOTx-NoVM protocol. Dub- $D_{ub}^k ~\forall k$, Trtt- $T_{rtt}^k~ \forall k$}
	\label{result}
\end{figure*} 
\section{Results and discussions}  
In this section, we evaluate the performance of the proposed EOTx-NoVM protocol in terms of the percentage of energy savings ($E_s$) and the percentage of DS traffic that violates the delay bound ($P_v$). Here, we define $E_s$ as the percentage of time the OLT transmitters are switched off while $P_v$ is defined as the percentage of the DS traffic, enters into the buffers at the OLT, that violates the delay bound ($D_{ds}^ k$). Firstly, we demonstrate the improvement in energy efficiency that is achieved by employing the proposed EOTx-NoVM protocol. Thereafter, we discuss how $E_s$ and $P_v$ change with the modification of its different associated parameters: delay bound, number of ONUs, round-trip time and sleep-to-wake-up. The performances are evaluated through simulations, performed in OMNET++ for a network run-time of $200$ s and all results are plotted with $95$\% of confidence interval. 
%The bandwidth of the feeder fiber and the access fiber are considered to be of 1 Gbps and 100 Mbps. 
Both the US traffic arrival to every ONUs and the DS traffic arrival to the OLT are assumed to be self-similar. The self-similar traffic is generated by aggregating $32$ ON-OFF Pareto sources \cite{ipact}. The shape parameters of ON and OFF periods of all ON-OFF Pareto sources are considered to be $1.2$ and $1.4$ respectively. All packets are assumed to be of the same size: $1500$ Bytes. The size of all buffers of  the OLT and ONUs are assumed to be $10$ Mb. 
%In all results, we consider the traffic load for the US and the DS traffic is the same.   
The tunning time between wavelengths $i$ and $j$ is considered to be $|i-j|T_t$ where $T_t$ is a constant \cite{twdm8}. All other considered parameters are enlisted in Table II. 
\subsection{Comparison with existing protocols}
Here, we compare the energy efficiency of the proposed EOTx-NoVM protocol with our previously proposed EO-NoVM protocol \cite{twdm8}, which is the only existing protocol for saving energy at the OLT transmitters to the best of our knowledge. To do so, in Fig. \ref{compari}, we plot the energy efficiency as a function of traffic load (in term of utilization factor) of both EOTx-NoVM and EO-NoVM protocols for $T_{sw}= 1$ ms and $2$ ms ($\forall k$) where $T_{sw}$ denotes the sleep-to-wake-up time of the OLT transmitters. While changing the Utilization Factor (UF), UF for both US and DS traffic are kept equal and this assumption is kept valid for all future results. The number of ONUs ($N$), number of wavelengths ($W$), delay bound for both US and DS traffic ($D_{ub}^ k$~$\forall k$), and round trip time ($T_{rtt}^ k$ ~$\forall k$) are considered to be $16$, $2$, $15$ ms, and $0.2$ ms respectively. It can be observed from Fig. \ref{compari} that a significant improvement in energy efficiency (up to $\sim 45\%$) can be achieved by employing the EOTx-NoVM protocol as compared to the EO-NoVM protocol. This improvement is due to the fact that the EOTx-NoVM protocol exploits the opportunity of filling-up multiple voids completely for reducing the number voids which was unexploited in EO-NoVM algorithm. Since, at low load, when the grant size is very small, the opportunity of filling-up multiple voids completely is limited, the energy efficiency of the EO-NoVM and the EOTx-NoVM protocols are almost the same. However, if the load is sufficiently high then this opportunity of completely filling-up multiple voids is significantly high, causing a significant increase in energy efficiency as seen from Fig. \ref{compari}. 
\subsection{Performance Evaluation}
\begin{figure}[t]
  	\centering
  	\includegraphics[scale=.7]{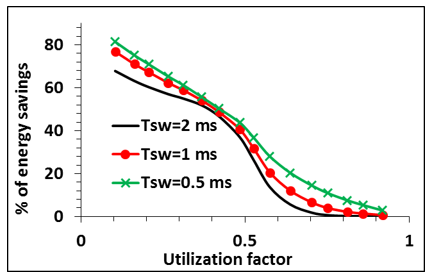}
  	\caption{Energy efficiency for different value of $T_{sw}$.}
  	\label{result1}
  \end{figure}

Here, we demonstrate the effect of $D_{ub}^k$, $N$, $T_{rtt}^k$ and $T_{sw}$ on $E_s$ and $P_v$ of the EOTx-NoVM protocol. To do so, in Fig. \ref{result} (a), (b) and (c), we plot the percentage of energy savings  as a function of traffic load (in term of utilization factor) for different values of $D_{ub}^k$ ($=10$ ms, $15$ ms, $20$ ms $\forall k$), $N$ ($=16$, $32$, $64$), and $T_{rtt}^k$ ($=0.2$ ms, $0.5$ ms, $1$ ms $\forall k$) respectively. Moreover, in Fig. \ref{result} (d), (e), and (f), we plot the percentage of DS traffic, whose delay bound gets violated (i.e., $P_v$) for the same values $T_{ub}^k$, $N$, and $T_{rtt}^k$ as mentioned above. It is evident that $P_v$ does not get affected by $T_{sw}$. Therefore, in Fig. \ref{result1}, we plot the percentage of energy savings vs utilization factor for different value of $T_{sw}$ ($=0.5$ ms, $1$ ms, $2$ ms). In all results, we consider that the value of $D_{ub}^k$ ($\forall k$), $N$, and $T_{rtt}^k$ ($\forall k$) are $10$ ms, 16, and $0.2$ ms respectively, unless they are explicitly mentioned.
\subsubsection{Effect of $D_{ub}^k$}
 As mentioned in Section \ref{ssec:outline}, in EOTx-NoVM, scheduling is performed in between $t_{lb}^{k,j}$ and $t_{ub}^{k}$. An increase in $D_{ub}^k$, increases the value of $t_{ub}^k$ (refer eq. (\ref{ub})) while keeping $t_{lb}^{k,j}$ unchanged (refer eq. (\ref{lb})). Thus, the increment of $D_{ub}^k$ increases the time interval over which scheduling can be performed, causing better opportunity of reducing the number of voids. This results in an increment of energy efficiency which can be seen from Fig. \ref{result}(a). Moreover, as the increment of $D_{ub}^k$ provides a longer time interval for scheduling, the possibility of satisfying the delay bound is more or in other word, $P_v$ gets reduced which can be seen from Fig. \ref{result}(d).   
 \subsubsection{Effect of $N$}
 In EOTx-NoVM, the DS traffic of ONUs are scheduled at every REPORT arrival from them. There exists a possibility of creation of a void, whenever scheduling of DS traffic is performed. An increment of the number of ONUs, increases the number of REPORT arrivals, causing an increases in the possibility of void creation. Hence, both the energy efficiency and the possibility of satisfying the delay bound decrease with an increase in the number of ONUs as seen from Fig. \ref{result}(b) and \ref{result}(e) respectively. Further, Fig. \ref{result}(b) depicts that this decrement of energy efficiency is very small and hence, the EOTx-NoVM protocol can provided a significant amount of energy efficiency even if large number of ONUs are present.       
 \subsubsection{Effect of $T_{rtt}^k$}
 In EOTx-NoVM, the US traffic is scheduled by using the previously proposed EO-NoVM protocol \cite{twdm8}. We have also demonstrated in \cite{twdm8} that the energy efficiency reduces with an increase in $T_{rtt}^k$ which implies more number of voids are created. Moreover, in EO-NoVM, if a new void is created then the US traffic is scheduled as late as possible. Thus, the time interval between two consecutive REPORT messages increases with an increase in $T_{rtt}^k$. As a result, the size of the void is quite large and hence, the possibility of complete fill-up of voids reduces especially at low load when the grant size is very low. Thus, at low load, the increment of $T_{rtt}^k$ reduces energy efficiency as seen from Fig. \ref{result}(c). However, at high load, grant size being very large, this effect becomes insignificant. Thus, the effect of changing the value of $T_{rtt}^k$ on both energy efficiency and the delay violation is limited as seen from Fig. \ref{result}(c) and Fig. \ref{result}(f) respectively.   
 \subsubsection{Effect of $T_{sw}$}
 Whenever the OLT switch off its transmitters, $T_{sw}$ duration is required to switched them on. In EOTx-NoVM, the transmitters are switched off for saving energy at all voids, whose size is more than $T_{sw}$. Note that the scheduling is unaffected by $T_{sw}$. Thus, when $T_{sw}$ decreases, the number of voids remain the same while in every void, the transmitters can be switched off for a longer duration resulting in an increment of energy efficiency. 
\section{conclusion}
  In this paper, we design a protocol, namely EOTx-NoVM, for saving energy at the OLT transmitters in TWDM-EPON by minimizing the number of voids while satisfying the SLAs. Simulation results depict that a significant improvement in energy efficiency (up to $\sim 45\%$) can be achieved by employing the EOTx-NoVM protocol when compared with existing protocols. We have also demonstrated that the probability of delay bound violation is quite low even at high load. For example, for the case 16 ONUs and 2 wavelengths, the probability of delay bound violation of the DS traffic is lower than $1\%$ even up to the utilization factor of $0.8$. The reduction in energy efficiency with the increment of the number of ONUs and round-trip time (after a certain load) is insignificant. Thus, EOTx-NoVM protocol is equally useful if the network is scaled up or used for long-reach EPON. Reduction in sleep-to-wake-up time increases energy efficiency drastically which promotes research on designing advance circularities for reducing the sleep-to-wake-up time of OLT transmitters. Consideration of different class of services and the interdependency of the protocols for saving energy at the OLT and ONUs are the future scope of the paper.  
    
\bibliographystyle{IEEEtran}
	
\end{document}